\documentstyle[11pt,aaspp4]{article}
\newcommand\beq{\begin{equation}}
\newcommand\eeq{\end{equation}}

\begin{document}
\title{Consequences of Interstellar Absorption for Models of
Anomalous X-ray Pulsars}
\author{Rosalba Perna$^{1,3}$, Jeremy Heyl$^2$ \& Lars Hernquist$^3$}
\medskip
\altaffiltext{1}{Harvard Junior Fellow}
\altaffiltext{2}{Lee A. DuBridge Postdoctoral Scholar, Theoretical 
Astrophysics, California Institute of Technology, Pasadena, CA 91125}
\altaffiltext{3}{Harvard-Smithsonian Center for Astrophysics, 60 Garden Street,
Cambridge, MA 02138}

\begin{abstract}

We examine properties of thermal radiation emitted by strongly
magnetized neutron stars (NSs). In particular, we show that the
pulsation amplitudes of the energy-integrated flux are an increasing
function of the absorption column density to the source. This is
especially important for the interpretation of the Anomalous $X$-ray
Pulsars (AXPs) as cooling neutron stars with high magnetic fields. We
show that the high-pulsation amplitudes observed in these objects are
consistent with cooling models, if the large amount of absorption to
these sources is taken into account. We also show that cooling models
imply inferred radii of the emitting regions on the order of $\sim
5-6$ times smaller than the actual NS radii, again in agreement with
observations. 

\end{abstract}


\section{Introduction}

In the past few years, $X$-ray observations have revealed a new class
of $X$-ray sources, the so-called anomalous $X$-ray pulsars (AXPs)
(Mereghetti \& Stella 1995; van Paradijs, Taam \& van den Heuvel
1995).  AXPs are sources of pulsed $X$-ray emission, with persistent
luminosities $L_x\sim 10^{35} - 10^{36}$ erg ${\rm s}^{-1}$ and soft
spectra. Their periods lie in a very narrow range, between 6 and 12
seconds, and their characteristic ages are of order $10^3 - 10^5$
years.  Their distances are usually very large, as inferred from
the typically large column densities to them ($N_{\rm H}\sim 10^{22}$
cm$^{-2}$).

Due to the long period of these objects, it is clear that their
$X$-ray luminosity cannot be rotation-powered. Two broad classes of
models have emerged to explain their $X$-ray emission. In one, the
radiation is assumed to be powered by accretion from binary companions
of very low mass (Mereghetti \& Stella 1995), from the debris of a
disrupted high mass companion (van Paradijs, Taam \& van den Heuvel
1995; Ghosh, Angelini \& White 1997), from the interstellar medium
(Wang 1997), or from the debris falling back after the supernova
explosion (Chatterjee, Hernquist \& Narayan 2000; Alpar 1999, 2000;
Marsden et al. 1999). In the second class of models, AXPs are
hypothesized to be isolated, ultramagnetized neutron stars, with field
strengths in the range $10^{14} - 10^{15}$~G, a.k.a. ``magnetars''
(Duncan \& Thompson 1992).  The $X$-ray luminosity could then be
powered by either magnetic field decay (Thompson \& Duncan 1996; Heyl
\& Kulkarni 1998) or by residual thermal energy (Heyl \& Hernquist
1997a,b).

There are several observational constraints that can be used to
discriminate among various models. The presence of an optical
counterpart and observable Doppler shifts in the frequency of the
$X$-ray pulses would suggest the presence of an optical counterpart,
but this has not been observed so far.  Optical and longer wavelength
emission (Hulleman et al 2000; Perna, Hernquist \& Narayan 2000) would
indicate the presence of an accreting fossil disk.  Any viable model,
however, must also be consistent with the observed high pulse
fractions and small emitting areas inferred for the $X$-ray 
radiation from these objects. 

In this {\em Letter} we examine whether the
observed pulse fractions and inferred emitting
areas are consistent with a model for AXPs in which their $X$-ray
emission is powered by thermal radiation from highly magnetized
neutron stars.  We show that the amplitudes of the pulsations
for the total flux observed in a given band are strongly sensitive to
the amount of absorption to the source, and, more specifically,
increase with it.  To understand this point, one first needs to notice
that, for typical local blackbody (BB) emission, the energy-dependent pulse
fractions increase with energy, as a consequence of the fact that the
hardness of the blackbody spectrum becomes larger at higher energies. 
This point was well elucidated by Page (1995) and Page \& Sarmiento (1996):
the ratio of the BB fluxes emitted at energy $E$ by two regions of areas
$A_1$ and $A_2$ with respective temperatures $T_1$ and $T_2$ is
\beq
\frac{F_{\rm BB}(E,T_1)}{F_{\rm BB}(E,T_2)}=\frac{A_1}{A_2}
\frac{\exp(E/kT_2)-1}{\exp(E/kT_1)-1}\;,
\label{eq:bb}
\eeq 
which is an increasing function of $E$ if $T_1> T_2$.  This
property remains valid for atmospheric models with beaming $\propto
(\cos\theta)^n$, and, furthermore, the increase of the pulse fractions
with energy becomes stronger with increasing $n$.  

Absorption by the intervening interstellar medium (ISM) does not
affect the energy dependent pulse fractions (though this is rigorously
true only for a perfect detector), as photons emitted at a given
energy are reduced by the same amount independently of the 
phase of the rotation. However, pulse fractions measured from flux which is
integrated over a given band do depend on absorption.  Absorption, in fact,
reduces the low-energy part of the flux with respect to the
high-energy one. As a result, the main contribution to the pulse
fractions is weighed more and more by the high energy component as the
amount of absorption is increased.  

We compute the expected pulse fractions and inferred areas for typical
parameters of AXPs and a thermal cooling model for their emission,
which includes processing through an atmosphere.  
We show that the inferred properties are consistent with
observations.

\section{Model}

Our model is that of a neutron star (NS) cooling through an accreted
envelope and with an intense, dipolar magnetic field. Heyl \& Hernquist
(1998b) showed that, for $B_p\ga 10^{14}$ G, the flux transmitted
through the envelope is well described by $F\propto \cos^2\psi$, where
$\psi$ is the angle between the radial direction and the magnetic
field.  For a dipolar field,
$\cos^2\psi=4\cos^2\theta_p/(3\cos^2\theta_p+1)$ (Greenstein \& Hartke 1983),
with $\theta_p$ being the magnetic colatitude (angle between radial
direction and magnetic pole). It is given by 
\beq
\cos\theta_p=\cos\theta\cos\alpha+\sin\theta\sin\alpha\cos\phi\;,
\label{eq:costeta}
\eeq 
where $\alpha$ is the angle that the magnetic pole makes with the
line of sight, and $(\theta,\phi)$ are the usual colatitude and
longitude in spherical coordinates.  In our calculation, we assume
$\alpha=90^\circ$ (orthogonal rotator). Therefore our results on the
pulsation amplitudes must be interpretated as upper limits.  We assume
that the local emission, $n(E,T)$, is a blackbody spectrum modified by
the presence of an atmosphere, for which we adopt the semianalytical
model of Heyl \& Hernquist (1998a), but with the inclusion of limb
darkening
\footnote{The model is that of an accreted atmosphere with a
magnetized iron envelope, but note that the composition of the
envelope does not really affect any of the points made here.}.  This
leads to a beaming of the radiation $\propto \cos\delta$, with
$\delta$ being the angle between the normal to the NS surface and the
direction of the photon trajectory.  We will also consider cases with
a more intense beaming $\propto (\cos\delta)^n$, but not extreme
(i.e. we restrict our attention to $n\le 2$).  Static general
relativistic effects (i.e. light deflection and gravitational
redshift) are fully taken into account in our calculations.

Let us define $e^{-\Lambda_s}=\sqrt{1-R/R_s}$, where $R$ is the
radius of the NS star, and $R_s=2GM/c^2$ its Schwarzschild radius.
Here we assume $M=1.4 M_\odot$, and consider the range of radii
$2\le (R/R_s)\le 4$, all compatible with the currently available models for the
NS equation of state.  
Let $D$  be the distance from the star to the observer, and 
$N_{\rm H}$ the intervening column density.  
The flux measured by an observer at infinity is then given by 
\beq
f(E;N_{\rm H})=\frac{\pi R_\infty^2\;\sigma T^4_{p,\infty}}
{4\pi D^2}\frac{1}{k T_{p,\infty}}e^{-\sigma(E)N_{\rm H}}
\int_0^1 2xdx\int_0^{2\pi}
\frac{d\phi}{2\pi}\; I_0(\theta,\phi) \;n[Ee^{-\Lambda_s};T_{s}(\theta,\phi)]\;,
\label{eq:flux}
\eeq in units of phot cm$^{-2}$ s$^{-1}$ keV$^{-1}$.  Here $x=\sin\delta$,
$R_\infty\equiv Re^{\Lambda_s}$, and $T_{p,\infty}\equiv T_{p}e^{-\Lambda_s}$, where
$T_p$ is the temperature at the pole. 
The general relativistic effects of light deflection are taken into
account through the ray-tracing function $\theta=\theta(x)$, as
described in Page (1995). The function $I_0(\theta,\phi)$ sets the
normalization of the flux at each point $(\theta,\phi)$ of the NS surface 
\beq
I_0(\theta,\phi)=\frac{4\cos^2\theta_p}{3\cos^2\theta_p+1}\;
(\cos\theta_p)^{0.4}\;,
\label{eq:I0}
\eeq 
where, following Heyl \& Hernquist (1998b), we have assumed a
further dependence of the flux on $B^{0.4}$. The local temperature on the star is
given by
$T_{s}(\theta,\phi)=T_p[I_0(\theta,\phi)]^{1/4}$, while
for the photoabsorption cross section $\sigma(E)$ we use the
analytical fit provided by Morrison \& McCammon (1988).

Let now $E_{\rm min}$ and $E_{\rm max}$ define the boundaries of the 
band of observation, and let us define 
$F(N_{\rm H})=\int_{E_{\rm min}}^{E_{\rm max}}dE f(E;N_{\rm H})$. Then,
if the column density to the source is $N_{\rm H}$, the pulse fraction in 
that given band is
\beq
pf(N_{\rm H})=\frac{F^{\rm max}(N_{\rm H})-F^{\rm min}(N_{\rm H})}
{F^{\rm max}(N_{\rm H})+F^{\rm min}(N_{\rm H})}.
\label{eq:pf}
\eeq

In Figure 1, we plot this function for fixed values of $T_p\equiv T_{eff}(\psi=0)$
and $n$, while exploring the dependence on $R/R_s$.  Our results agree with those
of DeDeo, Psaltis \& Narayan (2000) in the limit of $N_{\rm H}\rightarrow 0$. 
However, for large column densities, our plots show that the pulsation amplitudes
can significantly increase. Therefore, neglecting this effect can lead to
misleading interpretations.

In Figure 2, we consider a typical range of column densities inferred for AXPs
and show, for an intermediate value of $R/R_s$, the dependence of pulse fractions
on temperature and degree of beaming. We have considered a range of temperatures that
is typical for AXPs, if one accounts for the fact that processing through
the atmosphere typically results in an overestimate of the effective temperature
by a factor of $\sim 2-3$. Note that the pulsation amplitudes in a given energy band
increase as $T_p$ decreases. This is due to the fact that, as $T_p$ decreases,
the spectrum is shifted towards lower energies. This in turn leads, for the same amount
of absorption, to a larger suppression of softer photons, and hence to an increase
of the pulse fractions. Figure 2  clearly shows that a thermal
model for highly magnetized NSs can produce pulsation amplitudes which are
consistent with those observed so far (see Chakrabarty et al. 2000 for a summary). 

Another property that must be accounted for by any viable model for
AXPs is the inferred size of the emitting region.  To estimate the
effective emitting areas that an observer would deduce for an highly
magnetized, cooling neutron star as in our model, we generated a
spectrum using Equation (\ref{eq:flux}). We then used the software 
XSPEC to convolve it with the instrument response (we used {\em
CHANDRA}) and generate a set of simulated data.  These were then
fitted using a blackbody function. For a star of radius $R=\{2,3,4\}R_s$
(corresponding to $R=\{8.25,12.37,16.4\}$ km), we found, 
respectively\footnote{These values are not very dependent on $T_p$ and $n$,
within the range of values considered for the AXPs.}, 
$R_\infty\simeq$ \{1.3,2.3,3.5\} km, significantly smaller than the
stellar radius, but comparable to the parameters inferred for AXPs.

\section{Conclusions}

We have examined the properties of a model in which the $X$-ray
emission of AXPs is thermal radiation from highly magnetized cooling
neutron stars.  We have shown that the amplitudes of the pulsations in
a given energy band are an increasing function of the column density
to the source. If the typically high column densities inferred for
AXPs are taken into account, then even a moderate beaming can lead to
pulse fractions comparable to those observed so far.

We have used the software XSPEC to generate simulated data from our model
and then used these data to find the best-fit blackbody spectrum. We find
that the best fit implies an effective radius of the star which is typically
$\sim 5-6$ times smaller than the actual radius. This is again in
agreement with the radii of the emitting regions inferred from the
observations of AXPs.

Our results have clear implications for efforts to discern the nature
of a variety of X-ray point sources.  AXPs have variously been
interpreted either as accreting sources, or as isolated, strongly
magnetized neutron stars.  If interstellar absorption is not taken
into account, then the large observed pulse fractions appear to be
inconsistent with those predicted by thermal cooling models, as found
in recent work by DeDeo et al. (2000).  However, as we have demonstrated,
interstellar absorption is able to boost the pulsed fraction measured
by a distant observer relative to that which would be seen otherwise
to the point that a thermal cooling model can easily accommodate the
pulsed fractions associated with observed AXPs, if the high inferred
column densities to these sources are properly taken into account.
Consequently, we do not believe that pulsed fractions discriminate
between thermal models for AXP emission from those on which the X-ray
luminosity is accretion-powered.

Likewise, interstellar absorption complicates efforts to constrain the
sizes of thermally emitting areas on the surfaces of neutron stars.
The spectrum of the X-ray point source discovered at the center of the
Cas A supernova remnant (Tannanbaum 1999) has now been discussed
separately by a number of groups (Pavlov et al. 1999; Umeda et al.
1999; Chakrabarty et al. 2000).  Blackbody fits to the observed
spectra yield emitting areas that are only a small fraction of a
neutron star surface.  These results cannot be explained by other
models which interpret the spectra using hydrogen atmospheres, but
which ignore both the non-uniform temperature distribution expected
for a strongly magnetized neutron star and the consequences of
interstellar absorption.  As we have indicated, when all these various
effects are accounted for, there does not appear to be a discrepancy
between the effective emitting area of a strongly magnetized neutron
star and the values which have been derived for objects like Cas A
and the AXPs.

\begin{figure}[t]
\centerline{\epsfysize=5.7in\epsffile{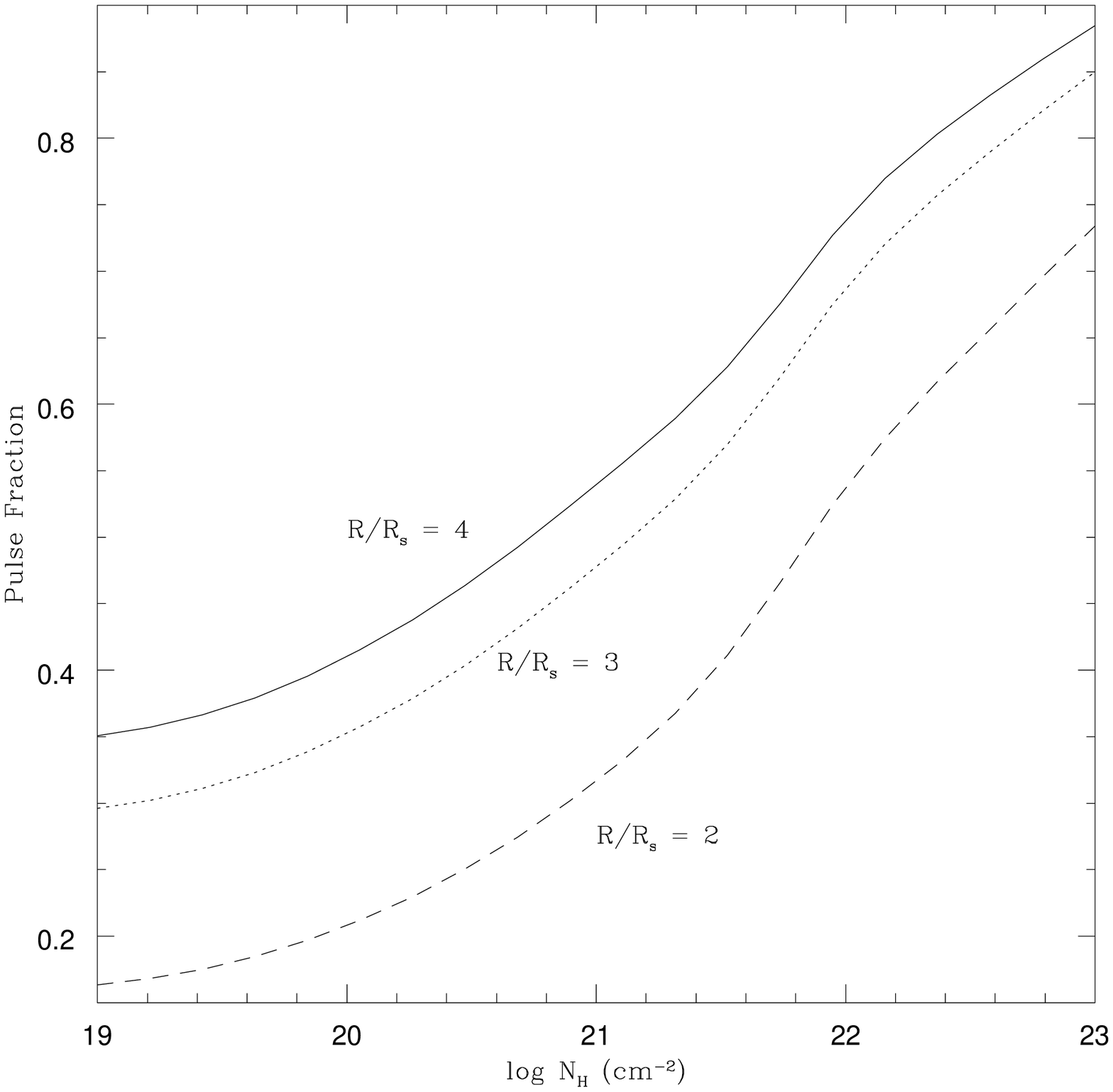}}
\caption{Dependence of the pulsation amplitudes on the absorption
column density to the source, for different values of the NS radius.
The flux here is assumed to be observed in the (0.1-10) KeV band, the
temperature at the magnetic pole 
is $T_p=10^6$ K, and the degree of beaming is $n=2$.
It can be seen that a large amount of absorption can significantly
increase the observed pulse fractions.}
\label{fig:1}
\end{figure}

\begin{figure}[t]
\centerline{\epsfysize=5.7in\epsffile{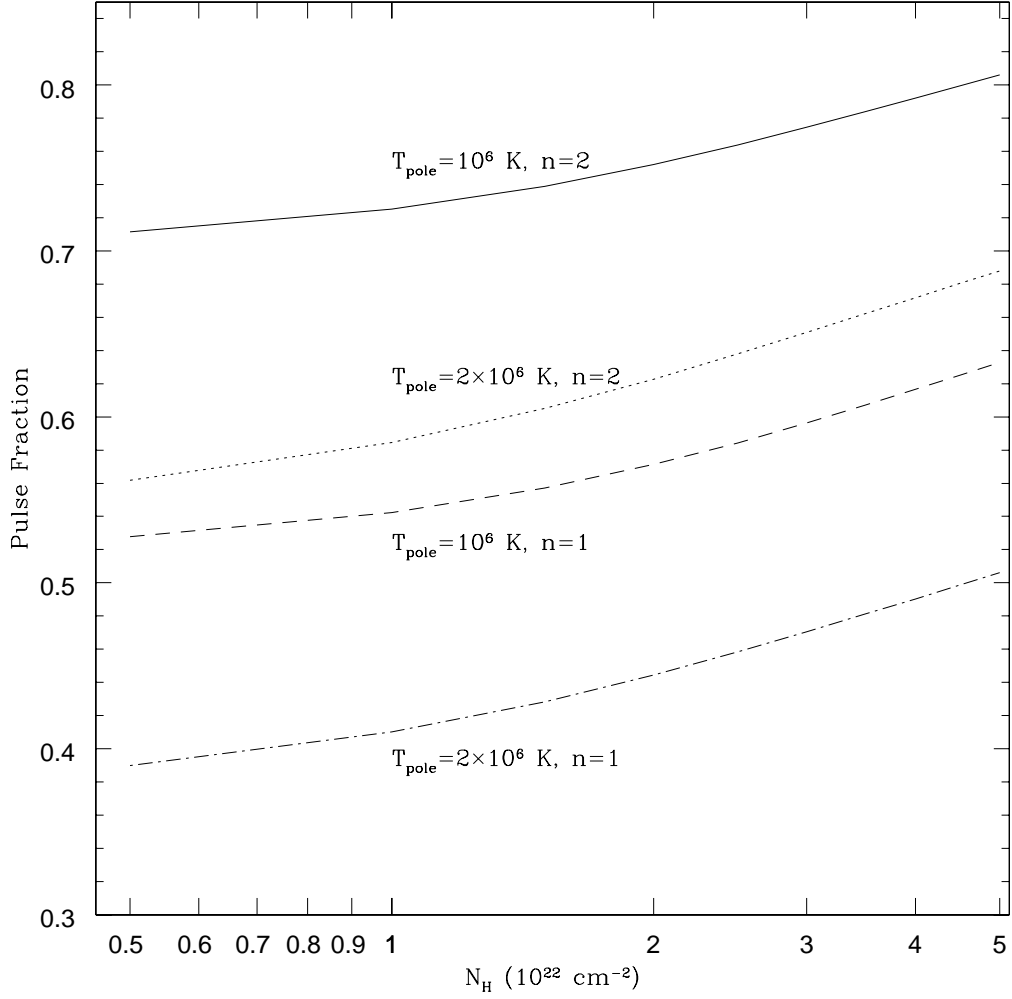}}
\caption{Pulsation amplitudes in the range of temperatures and column
densities corresponding to those typically inferred for AXPs. The flux
is taken in the (1-10) KeV band, and the NS radius is assumed to be
$R=3 R_s$.  These values of the pulse fractions are consistent with
those observed.}
\label{fig:2}
\end{figure}

\end{document}